\documentclass[prd, preprint, showpacs, showkeys, nofootinbib]{revtex4}    
\usepackage{amssymb}
\usepackage{amsmath}
\usepackage{graphicx}
\usepackage[dvips]{epsfig}
\usepackage[usenames,dvipsnames]{color}
\usepackage{hyperref}

\hypersetup{
    colorlinks=true,         
    linkcolor=blue,          
    citecolor=red,        
    urlcolor=Violet             
}

\newcommand{\eq}[1]{(\ref{eq:#1})}

\newcommand{\scn}[1]{Sec.~(\ref{sec:#1})}

\newcommand{\diby}[2]{\ensuremath{\frac{\partial #1}{\partial #2}}}
\newcommand{\equa}[1]{\begin{equation} #1 \end{equation}}
\newcommand{\pb}[2]{\ensuremath{\lf\{#1,#2 \rt\}}}
\newcommand{\myexp}[1]{\ensuremath{\exp\lf\{ #1 \rt\}}}

\def\lf {\ensuremath{\left}}
\def\rt {\ensuremath{\right}}
\def\ra {\ensuremath{\rightarrow}}

\def\qand {\ensuremath{\quad\text{and}}}

\newcommand{\ta}[2]{\ensuremath{\lf.t_\alpha\rt.^{#1}_{#2}}}

\def\pia {\ensuremath{\pi_\alpha}}
\def\wa {\ensuremath{\omega^\alpha}}
\def\lin {\ensuremath{\mathcal{H}_\alpha}}
\def\diff {\ensuremath{\mathcal{H}^a}}
\def\diffadm {\ensuremath{\mathcal{H}^a_\text{ADM}}}
\def\ham {\ensuremath{\mathcal{H}}}
\def\hamadm {\ensuremath{\mathcal{H_\text{ADM}}}}
\def\Na {\ensuremath{N^\alpha}}

\def\acs {\ensuremath{\mathcal{A}}}
\def\covd {\ensuremath{\mathcal{D}_\omega}}

\def\dw {\ensuremath{\dot{\omega}^\alpha}}
\def\htot {\ensuremath{H_\text{T}}}
\def\dq {\ensuremath{\dot {q}}}
\def\dxi {\ensuremath{\mathcal{D}_\xi}}

\begin{document}

\title{A Definition of Background Independence}
\date{\today}
\author{Sean Gryb}
\affiliation{Perimeter Institute for Theoretical Physics\\Waterloo, Ontario N2L 2Y5, Canada}
\affiliation{Department of Physics and Astronomy, University of Waterloo\\Waterloo, Ontario N2L 3G1, Canada}
\email{sgryb@perimeterinstitute.ca}

\begin{abstract}
    We propose a definition for background (in)/dependence in dynamical theories of the evolution of configurations that have a continuous symmetry and test this definition on particle models and on gravity. Our definition draws from Barbour's best--matching framework developed for the purpose of implementing spatial and temporal relationalism. Among other interesting theories, general relativity can be derived within this framework in novel ways. We study the detailed canonical structure of a wide range of best matching theories and show that their actions must have a local gauge symmetry. When gauge theory is derived in this way, we obtain at the same time a conceptual framework for distinguishing between background dependent and independent theories. Gauge invariant observables satisfying Kucha\v r's criterion are identified and, in simple cases, explicitly computed. We propose a procedure for inserting a global background time into temporally relational theories. Interestingly, using this procedure in general relativity leads to unimodular gravity.
\end{abstract}

\maketitle
\tableofcontents


\section{Introduction}

``Background independence'' is a term so often used (and misused) in the quantum gravity literature that I won't even begin to attempt to give a comprehensive list of citations to support this claim. The statements made on this subject are as extensive as they are subtle and tend to vary significantly between fields and individuals. For this reason, I will \emph{not} attempt to consolidate these statements into one coherent picture. Rather, I will provide a concrete proposal that, in a specific context, is successful at distinguishing particular examples of theories generally understood to be either background dependent or independent. To accomplish this, I will study the best--matching framework, developed by Barbour and collaborators in \cite{barbour:nature,barbourbertotti:mach,barbour:eot,barbour:scale_inv_particles,barbour_el_al:scale_inv_gravity,barbour_el_al:rel_wo_rel,anderson:rel_wo_rel_vec,barbour_el_al:physical_dof,Anderson:rwor2}, by performing a detailed canonical analysis of a general class of models. Examples of models treated in the cited papers using best matching include Newtonian particle mechanics, Maxwell theory, and general relativity.


In best matching, the variational principle used is non--standard in its use of certain auxiliary fields (to be defined later) used to make the theory satisfy Poincar\' e's principle -- a principle proposed by Barbour to implement Mach's principle. We will discuss Poincar\' e's principle and its relation to Mach's principle later in more detail later. Then, we shall show how best matching leads to an alternative approach to the local gauge principle. From the point of view of this paper, one advantage of this approach is that it automatically provides a framework for distinguishing between background dependence and independence.

So far, the framework -- including the definition of the Machian variational principle used -- has been developed almost entirely in the Lagragian picture with only a cursory mention of the canonical formalism. In this paper, we develop a detailed canonical analysis of the best--matching framework and propose a canonical version of the Machian variational principle used for the auxiliary fields. There are several benefits to working out the details of the canonical framework. It allows us to deduce all of the gauge transformations of the theory, gives us a formal definition for the gauge--independent observables, and paves the way for the canonical quantization. When best matching is applied to Newtonian particle mechanics, the gauge--independent observables can be explicitly computed. For these reasons, I believe that the canonical formalism helps provide a solid structural backbone to a framework that has been built on an exceptional conceptual foundation.

One important issue that the best--matching framework can shed light on is that of background independence.\footnote{For a discussion of these issues in the spirit presented here, see \cite{Barbour:general_cov_bm}.} We will show that best matching suggests a natural and precise definition of a \emph{background}. By this definition, background independent and dependent theories are differentiated by  the rules of variation of the auxiliary fields. Specifically, we find it convenient to associate a background to a particular continuous symmetry of the configurations. Then, our definition implies the following: if the theory places ``physical meaning'' (defined more precisely later) on the location of a system along a symmetry direction then it has a background with respect to this symmetry. If it does not, then it is background independent with respect to this symmetry. Under this definition, Newtonian particle mechanics is background \emph{dependent} with respect to rotations (it places absolute meaning to the absolute orientation of the system) while general relativity is background \emph{independent} with respect to diffeomorphisms. This definition both allows us to distinguish between theories that are simply ``covariantized'' and those that are truly background independent and to take a background independent theory and make it background dependent (or vice-versa).

When applying this rationale to theories invariant under time reparameterizations, it is convenient to distinguish between two geodesic principles. The first: Jacobi's, is a square root action that is manifestly time independent. The second is a parametrized version of Hamilton's principle. It can be made either background dependent, in which case it expresses a Newtonian absolute time, or background independent, in which case it is equivalent to Jacobi's principle. As we will see, testing this procedure in geometrodynamics leads either to general relativity, when the theory is kept independent of a background time, or to unimodular gravity, when a background time is introduced.

\subsection{Structure of the Paper}

In this paper, I try to treat the widest class of models possible. For this reason, I consider arbitrary configuration spaces and symmetry groups and work out several concrete examples. However, there exists a natural division between finite--dimensional and infinite--dimensional models that is reflected in the structure of the paper. The simplicity of the finite--dimensional case allows for explicit solutions that lead to concrete statements about the structure of the theory. In particular, gauge--independent observables can be identified. With these concrete results, I motivate my definition of background independence. Despite the simplicity of the finite--dimensional case, many useful models can be treated within this framework. These include non--relativistic and relativistic particle models and cosmological models like mini--superspace. More interesting still are the infinite--dimensional models, which include geometrodynamics, even though less can be done in terms of explicit calculations. Nevertheless, the definition of background independence we are led to can be used to insert a background time into general relativity and leads to unimodular gravity.

The logic of the paper is as follows: the detailed structure of the finite--dimensional models is worked out and these results are used to motivate a definition of background independence. Then, the basic canonical structure of the infinite--dimensional models is given and it is shown how this definition can be used in a simple case.

\section{Finite Dimensional Relational Models}

In this section we develop the canonical structure of best matching and use it to study finite--dimensional models whose configurations have continuous symmetries. We first descirbe a generalized formulation of Jacobi's principle, which implements temporal relationalism, then best matching, which implements spatial relationalism.


\subsection{Relational Mechanics Using Jacobi's Principle and Best Matching}

\subsubsection{Jacobi's Principle}\label{sec:jacobi_principle}

Jacobi's principle is a geodesic principle on a configuration space, $\acs$. It is expressed by the action
\equa{\label{eq:jacobi_gen}
	S_J = \int_{q_\text{in}}^{q_\text{fin}} d\lambda\, \sqrt{g_{ab}(q) \dot{q}^a(\lambda)\dot{q}^b(\lambda)}.
}
The $q$'s are the configuration space variables and $g_{ab}$ is a metric that is a function only of $q$ (and not of its $\lambda$-derivatives). A dot represents differentiation with respect to $\lambda$. Given that $S_J$ is invariant under reparametrizations of $\lambda$, the parameter $\lambda$ is completely arbitrary. It has been written explicitly so that we can use it as an independent variable in the canonical analysis.

We will find it convenient to decompose the metric in terms of the conformal metric $\gamma_{ab} = g_{ab}/g$, where $g = \det g_{ab}$, and a conformal factor $e^\phi = g$ such that
\equa{\label{eq:conf_decomp}
	g_{ab} = e^{\phi(q)} \gamma_{ab}.
}
In many applications, the conformal factor of the metric on $\mathcal A$ is defined as the negative of twice the potential energy. When considering the dynamics of non--relativistic particles, the configuration space is just the space of particle positions $q^i$. The metric $g_{ab}$ leading to Newton's theory happens to be conformally flat so that
\equa{ 
	\gamma_{ab} = \eta_{ab},
}
where $\eta$ is the flat metric with Euclidean signature.\footnote{The units can be chosen so that all of elements of $\eta$ are 1. Particles with different masses can be considered by replacing $\eta$ with the suitable mass matrix for the system.} In general, the metric $g_{ab}$ is a specified (ie, \emph{non}--dynamical) function on $\acs$. 

For a more familiar form of Jacobi's principle, define $e^\phi \equiv -2V$ and $2T \equiv \eta_{ab} \dot{q}^a\dot{q}^b$, where $V\equiv V' - E$, $V'$ is the standard potential energy, $T$ is the kinetic energy, and $E$ is the total energy of the system (which has been absorbed into the definition of $V$). This leads to Jacobi's action for a non--relativistic system of particles\footnote{For an introductory treatement of Jacobi's principle see section V.6-7 of Lanczos's book \cite{lanczos:mechanics}.}
\equa{
	S_J = \int_{q_\text{in}}^{q_\text{fin}} d\lambda\, 2 \sqrt{T} \sqrt{-V}.
}
From now on, we will use the action (\ref{eq:jacobi_gen}), making use of the decomposition (\ref{eq:conf_decomp}) only when necessary. This allows us to work directly with geometric quantities on $\acs$.

Because the Jacobi action \eq{jacobi_gen} is the length of a path on configuration space, its variation will lead to the geodesic equation
\equa{
    \ddot q^a + \Gamma^a_{bc} \dot q^b \dot q^c = \kappa(\lambda) \dot q^a,
}
where $\kappa \equiv d \ln \sqrt{g_{ab} \dot{q}^a\dot{q}^b}/ d\lambda$ and $\Gamma^a_{bc} = \frac{1}{2} g^{ad} (g_{db,c} + g_{dc,b} - g_{bc,d})$ is the Levi-Civita connection on $\acs$.

The choice of the parameter $\lambda$ is important. Normally, one would like to set $\kappa = 0$ with an affine parameter. However, for metrics of the form \eq{conf_decomp} with $\gamma = \text{const}$, there is another special choice of $\lambda$ that simplifies the geodesic equation. If we choose the parameter $\tau$ such that
\equa{
    \frac{d\tau}{d\lambda} = \frac{\sqrt{g_{ab} \dot{q}^a\dot{q}^b}}{e^\phi},
}
the geodesic equation becomes
\equa{
    \gamma_{ab} \frac{d^2 q^b}{d\tau^2} = \frac{1}{2}\partial_a e^\phi.
}
In the case of non--relativistic particles, $e^\phi = -2V$ and $\gamma = \eta$ so that the geodesic equation is Newton's $2^{\text{nd}}$ law. With these choices, $\tau = \sqrt{-\frac{T}{V}}$ is Barbour and Bertotti's (BB's) \emph{ephemeris} time. On top of simplifying the equations of motion, the parameter $\tau$ has the amazing property that its projection onto isolated subsystems is \emph{equal} to its definition on that subsystem if one were to ignore the rest of the system. I take this as the mathematical statement of Barbour's \emph{marching in step} criterion \cite{Barbour:nature_of_time}. Because of this property, $\tau$ can be used to construct useful clocks that approximate the Newtonian time.

\subsubsection{Best Matching (Canonical Constraints)}

Best matching is a procedure first developed in \cite{barbourbertotti:mach} for implementing Mach's principle. In order to be able to use the procedure, one must first notice a continuous symmetry in the configurations of a physical system. There must exist a continuous group whose action on the configurations produces new configurations that are physically indistinguishable from the originals. In this paper, we will only consider the case where this symmetry is further reflected as a symmetry in the metric on $\acs$, although the more general case can also be treated (see \cite{barbour_el_al:physical_dof,barbour_el_al:scale_inv_gravity}). Once this symmetry is noticed, one introduces auxiliary fields whose role is to parametrize the symmetry. We will show that the presence of these auxiliary fields leads to primary first class constraints restricting the number of independent degrees of freedom of the system. The constrained system lives on a \emph{reduced configuration space} $\mathcal R$, which is equal to $\acs$ modded out by the symmetry group.

This reduction is crucial for implementing what Barbour calls \cite{barbourbertotti:mach} Poincar\' e's principle. This principle is based on the observation that, in the presence of a symmetry in the configurations, only independent data specifiable on the \emph{reduced} configuration space should affect the physical predictions of the theory. Thus, specifying initial conditions on $\acs$ gives more information than is necessary to evolve the system. As we will see, there is one extra piece of information for each symmetry. In the case where this symmetry is reflected in the metric, the extra information appears as a constant of motion. For a theory to be relational with respect to the symmetries, the physical predictions of the theory must not depend on such extra information. Poincar\' e's principle is then stated as follows: a relational theory must be determined uniquely by an initial point and direction\footnote{Only a direction is needed if Jacobi's principle is used to implement temporal relationalism.} in $\mathcal R$. Later, this will be a guiding principle for my definition of background independence.


We will now perform the canonical analysis of the best--matching procedure presented in \cite{barbour:scale_inv_particles}. The idea is to introduce the corrected coordinates
\equa{
    \bar{q}^a = G^a_b(\omega) q^b,
}
where $G^a_b(\omega) = \myexp{\omega^\alpha(\lambda) \ta{a}{b}}$ is an element of the group $\mathcal{G}$ generating the symmetries of the configurations (and, in our case, the metric $g$) and $\alpha$ ranges from $1$ to the dimension of the group. The group parameters, $\omega^\alpha$'s, are \emph{the} auxiliary fields of best matching and the $\ta{a}{b}$'s are generators of the local algebra. After defining the corrected coordinates, one replaces $q$ everywhere in the action with $\bar{q}$ then varies $\omega$ with a Mach variation.\footnote{This variation will be decribed in detail in \scn{mach_variation}.}

Take, for example, the case of non--relativistic particles. One might notice that all configurations of particles are symmetric under translations, rotations, and scale transformations. None of these operations will change \emph{anything} that can be measured by an observer inside the system. We could then use best matching to require that the dynamics reflect this symmetry. Choosing the generators
\begin{align}
    \text{Translations:} \qquad \ta{i}{j} &\ra \delta^i_j \partial_k\quad (\alpha\ra k = 1\hdots3) \label{eq:gen_trans}\\
    \text{Rotations:} \qquad \ta{i}{j} &\ra \epsilon_{ijl} q^l \partial_k\quad (\alpha\ra k = 1\hdots3) \\
    \text{Dilatations:} \qquad \ta{i}{j} &\ra \delta^i_j q^l \partial_l\quad (\alpha\ra 0 ) \label{eq:gen_dil}
\end{align}
we can \emph{best match} each of these symmetries.

In general, $\bar{q}$ is inserted into (\ref{eq:jacobi_gen}). Rearranging,
\equa{\label{eq:jacobi_bm}
    S_J(\bar{q}) = \int_\gamma d\lambda\, \sqrt{g_{ab}(\bar{q})\, G^a_c G^b_d\, \covd q^c \covd q^d},
}
where $\covd q^a = \dot{q}^a + \dot{\omega}^\alpha \ta{a}{b} q^b$ is the covariant derivative of $q$ with connection $\dot\omega$ along a trial curve $\gamma$ in $\acs$. Indeed, it can be thought of as the pullback onto $\gamma$ of a connection on the principal $\mathcal G$-bundle $\mathcal A$ over $\mathcal R$.

The action \eq{jacobi_bm} can be written in an illuminating form using the fact that our metric is symmetric under $\mathcal G$. The existence of global Killing vectors is expressed by the fact that the Lie derivative in the direction of the symmetry generators $\mathcal L_{t_\alpha q} g = 0$ is zero. Explicitly,
\equa{\label{eq:killing_loc}
    \ta{c}{\lf( a\rt.} g_{\lf. b\rt) c} + \partial_c g_{ab} \ta{c}{d} q^d = 0.
}
where the rounded brackets indicate symmetrization of the indices. This expression can be exponentiated to prove the following relation
\equa{\label{eq:killing_field}
    g_{ab}(\bar{q})\, G^a_c G^b_d = g_{cd}(q)
}
between the metric evaluated at the barred coordinates and the unbarred coordinates. Inserting this into \eq{jacobi_bm} gives
\equa{\label{eq:bm_jacobi_S}
    S_J(\bar{q}) = \int_{q_\text{in}}^{q_\text{fin}} d\lambda\, \sqrt{g_{ab}(q)\, \covd q^a \covd q^b}.
}

The $\omega$'s are varied using the Mach variation, discussed in \scn{mach_variation}, which brings them to their \emph{best--matched} values. The action (\ref{eq:bm_jacobi_S}) is that used in \cite{sg:ym_bm} to motivate the correspondence between best matching and gauge theory. These approaches are identical provided \eq{killing_field}, which is an expression of the global gauge invariance of the original action, is satisfied. Because we are dealing with gauge theories over configuration space and not the usual case over spacetime, \emph{global} gauge invariance refers to the invariance of the action under $\lambda$-independent group transformations of the coordinates. Best matching makes this global symmetry local in $\lambda$. Thus, it motivates the gauge principle. More generally, best matching can be extended to include actions that do not start out globally gauge invariant (see \cite{barbour_el_al:physical_dof,barbour_el_al:scale_inv_gravity}). The correspondence between these theories and standard gauge theories is still under investigation.


We can now proceed with the canonical analysis of the gauged Jacobi action (\ref{eq:bm_jacobi_S}). The momenta $p_a$, conjugate to $q^a$, and $\pia$, conjugate to $\omega^\alpha$, are
\begin{align}
    p_a &\equiv \diby{L}{\dot{q}^a} = \frac{g_{ab}\, \covd q^b}{\sqrt{g_{cd}\, \covd q^c \covd q^d }}, \label{eq:pa_jac} \qand \\
    \pia &\equiv \diby{L}{\dw} = \frac{g_{ab}\, \covd q^b}{\sqrt{g_{cd}\, \covd q^c \covd q^d }} \ta{a}{e} q^e. \label{eq:pia_jac}
\end{align}
It is easy to verify that these momenta obey the following primary constraints
\begin{align}\label{eq:jacobi_constraint}
    \ham &= g^{ab}\, p_a p_b -1 = 0, \qand \\
    \lin &= \pia - p_a \ta{a}{b} q^b = 0, \label{eq:lin}
\end{align}
where $g^{ab}$ is the inverse of $g_{ab}$. The quadratic \emph{scalar constraint} $\ham$ arises from the fact that $p_a$, according to (\ref{eq:pa_jac}), is a unit vector on configuration space. As a result, it gives a direction in $\acs$ only. $\ham$ reflects the irrelevance of the length of $\dot{q}$. The linear \emph{vector constraints} $\lin$ reflect the continuous symmetries of the configurations. They indicate that the phase space associated to $\acs$ contains equivalence classes of states generated by $\lin$. Later we will see that they are related to Noether's theorems. Note that $\ham$ and $\lin$ arise in very different ways. This seems to be reflected in the roles they play in the theory.

Using the fundamental Poisson Brackets (PBs)
\equa{
    \pb{q^a}{p_b} = \delta^a_b, \qand \quad \pb{\wa}{\pi_\beta} = \delta^\alpha_\beta,
}
we find that there are two sets of non--trivial PBs between the constraints. They are
\begin{align}
    \pb{\lin}{\ham_\beta} &= c_{\alpha\beta}^\gamma \ham_\gamma \label{eq:lie_algebra} \text{, and}\\
    \pb{\ham}{\lin} &= \partial_c g^{ab}\, p_a p_b \ta{c}{d} q^d - g^{ab} p_c p_{(a} \ta{c}{b)}. \label{eq:ham_lin_pb},
\end{align}
where $c_{\alpha\beta}^\gamma$ are the structure constants of the group. From \eq{lie_algebra}, we see that the closure of the vector constraints on themselves is guaranteed provided $\mathcal G$ is a Lie algebra. The PB's \eq{ham_lin_pb} vanish provided \eq{killing_loc} is satisfied. Thus, the closure of the constraints is guaranteed by the \emph{global} gauge invariance of the action.\footnote{In the more general context, the RHS of \eq{ham_lin_pb} could be treated as a secondary constraint introducing new auxiliary fields.} 

Because of the important role played by \eq{killing_loc}, it is illuminating to see the conditions under which \eq{killing_loc} is satisfied for particular models. In translationally invariant non--relativistic particle models the generators \eq{gen_trans} are used in \eq{killing_loc}. Being careful about particle and spatial indices (particle indices are labeled by $I$ and spatial indices are indicated by arrows) leads to the following condition on the potential
\equa{
    \sum_I \vec\nabla_I V = 0,
}
where $\vec\nabla_I = \diby{}{\vec q_I}$. This requires that the potential be translationally invariant. It is satisfied by potentials that are functions of the differences between the coordinates. The same argument applied to the rotations leads to a similar result: the potential must be rotationally invariant. The dilatations are different. They imply the following condition on the potential
\equa{\label{eq:dilatation_cc}
    \partial_c V\, q^c = -2V.
}
By Euler's theorem, this implies that the potential should be homogeneous of order $-2$ in $q^c$.

While the gauge invariance of the action is guaranteed for the rotations and translations by the gauge invariance of the potential, it is not for the dilatations. This is because the kinetic term has conformal weight $+2$ under global scale transformations of the $q$'s. Thus, the potential must have conformal weight $-2$ if the action is to be scale invariant. This is just the requirement \eq{dilatation_cc} and is equivalent to the \emph{consistency conditions} obtained in \cite{barbour:scale_inv_particles} but derived from different motivations and in the canonical formalism.

Finally, it is possible to work out the gauge transformations generated by the linear constraints $\lin$. Computing the PBs $\pb{q}{\lin}$ and $\pb{\wa}{\ham_\beta}$ we find $q$ and $\omega$ transform as
\begin{align}
    q^a &\ra e^{-\zeta^\alpha \ta{a}{b}} q^b \notag \\
    \wa &\ra \wa + \zeta^\alpha \label{eq:gauge_trans}
 \end{align}
under large gauge transformations parameterized by $\zeta^\alpha$. This is the \emph{banal invariance} noticed by Barbour in \cite{barbour:scale_inv_particles}. From the canonical analysis, this is a genuine gauge invariance of the theory. In standard gauge theory language, this corresponds to \emph{local} gauge invariance. However, because of the different nature of the connections used in this approach compared with Yang-Mills theory, \emph{local} in this context means local in $\lambda$ not local in spacetime. From the point of view of best matching, this local gauge invariance arises from the best matching procedure itself. It is not just the \emph{ad hoc} result of \emph{gauging} a global symmetry.



\subsubsection{Mach Variation} \label{sec:mach_variation}

Before computing the classical equations of motion and solving the constraints, we will describe the canonical Mach variation used for the auxiliary fields $\wa$. This non--standard variation plays a key role in our definition of background independence. For more details on the Lagrangian formulation of this variational principle, see 
\cite{barbour:scale_inv_particles} or \cite{barbour_el_al:scale_inv_gravity}.

The idea is that the auxiliary fields should be varied freely on the endpoints of \emph{any} interval along the trajectory. The implication of this free variation is that initial and final data cannot be specified for these variables. In this way, best matching implements Poincar\' e's principle.

To see how this works, we start with the canonical action:
\equa{\label{eq:omega_pi_action}
    S[q,p,\omega, \pi] = \int d\lambda \lf[ p\cdot\dot{q} + \dot\omega\cdot\pi - h(q,p, \omega, \pi) \rt].
}
We are concerned only with variations of the $\omega$'s and $\pi$'s since the $p$'s and $q$'s are treated as standard phase space variables. We need to determine the conditions under which the action will vanish if the $\omega$'s and the $\pi$'s are varied freely at the endpoints. The variation with respect to the $\pi$'s vanishes provided $\dot\omega = \diby{h}{p} = \pb{q}{h}$ regardless of the conditions on the endpoints. Thus, Hamilton's first equation is unchanged by the free endpoint condition. However, the procedure leading to Hamilton's second equation is modified.

After integration by parts, the variation of \eq{omega_pi_action} with respect to $\omega$ is
\equa{
    \delta_\omega S[q,p,\omega,\pi] = -\int d\lambda \lf[ \diby{h}{\omega} + \dot \pi \rt] \, \delta\omega + \lf. \pi\, \delta \omega \rt|_{\lambda_\text{in}}^{\lambda_\text{fin}} = 0.
}
The first term implies Hamilton's second equation
\equa{
  \dot{\pi} = -\diby{h}{\omega} = \pb{\pi}{h}.
}
However, because $\delta \omega$ is \emph{not} equal to zero on the endpoints, the second term will only vanish if $\pi(\lambda_\text{in}) = \pi(\lambda_\text{fin}) = 0$. This single free endpoint condition, however, is not enough. In order for the $\omega$ fields to be completely arbitrary, the solutions should be independent of where the endpoints are taken along the trajectory. This implies the \emph{Mach condition}, $\pi(\lambda) = 0$, \emph{everywhere}. The Mach condition ensures that the auxiliary fields are truly unphysical everywhere along the trajectory. It is an additional equation of motion. In Dirac's language, it is a \emph{weak} equation to be applied only \emph{after} taking Poisson brackets.

For metrics satisfying \eq{killing_field}, $\omega$ is a cyclic variable. This means that it enters the action only through its dependence on $\dot\omega$. In this case, $\pb{\pi}{h} = 0$ identically so that, by Hamilton's second equation, $\pi$ is a constant of motion. Normally, this constant of motion would be set by the initial and final data. The main effect of applying the Mach condition is to set this constant equal to zero, implementing Poincar\' e's principle.

\subsubsection{Classical Equations of Motion}\label{sec:part_jac_eom}

We are now in a position to compute the classical equations of motion of our theory. The definitions of the momenta, (\ref{eq:pa_jac}) and (\ref{eq:pia_jac}), imply that the canonical Hamiltonian vanishes, as it must for a reparametrization invariant theory. Thus, the total Hamiltonian $\htot$ is proportional to the constraints
\equa{
    \htot = N\ham + \Na\lin
}
where the \emph{lapse}, $N$, and \emph{shift}, $\Na$, are just Lagrange multipliers enforcing the scalar and vector constraints respectively. We use this terminology to emphasize that these Lagrange multipliers play the same role as the lapse and shift in general relativity.

The Mach variation implies
\begin{align}
    \dot{\omega}^\alpha &= \pb{\wa}{\htot} = \Na, \\
    \dot{\pi}_\alpha &= \pb{\pi}{\htot} = 0, \qand \\
    \pia &=0.
\end{align}
The $\wa$'s are seen to be genuinely arbitrary given that their derivatives are equal to the shift vectors. As expected, the $\pia$'s are found to be constants of motion set to zero by the Mach condition. Combining these results with the vector constraints (\ref{eq:lin}) requires that the generalized momenta associated to each symmetry vanish. In non--relativistic particle dynamics best matched under spatial translations, (\ref{eq:lin}) takes the form $\sum_I \vec p_I = 0$. This is the vanishing of the total linear momentum of the system. In the case of rotations, (\ref{eq:lin}) is the vanishing of total angular momentum of the system. Later, we will see that (\ref{eq:lin}) generalizes to the diffeomorphism constraint of general relativity.

There is an obvious connection to Noether's theorem. For actions invariant under the global symmetry condition \eq{killing_field}, the $\pia$'s are constants of motion and the linear constraints become a dynamically derived statement of the conservation of the Noether currents as obtained in Noether's first theorem. This is a result of parametrizing the symmetry using the corrected coordinates and making the $\omega^\alpha$ fields dynamical. The Mach condition requires, in addition, that the Noether charges vanish.

We now perform a standard variation of the $q$'s and $p$'s. A short calculation shows that Hamilton's first equation $\dot{q}^a = \pb{q^a}{\htot}$, can be re-written as
\equa{\label{eq:h1_pq}
    p_a = \frac{1}{2N} g_{ab}\, \lf.G^{-1}\rt.^b_c \diby{}{\lambda}\lf( G^c_d q^d \rt),
}
where we have made use of the definition $\lf.G^{-1}\rt.^a_b = \myexp{-\wa\ta{a}{b}}$. Note that $G$ can be rewritten in terms of the shift vectors using the equation of motion $\dw = \Na$. By (\ref{eq:killing_field}), we find that, in terms of barred quantities, (\ref{eq:h1_pq}) becomes
\equa{\label{eq:h1_pbar}
    \bar{p}_a = \frac{1}{2N} g_{ab}(\bar{q})\, \dot{\bar{q}}^b,
}
where $\bar{p}_a = \lf.G^{-1}\rt.^b_a p_b$.

Hamilton's second equation gives
\equa{
    \dot{p}_a = -N (\partial_a g^{bc}) p_b p_c + \Na p_b \ta{b}{a},
}
which, upon repeated use of (\ref{eq:killing_field}), leads to
\equa{
    \dot{\bar{p}}_a = -N (\bar{\partial}_a g^{bc}(\bar{q})) \bar{p}_b \bar{p}_c.
}
Thus, the equations of motion can now be written purely in terms of the barred quantities:
\equa{\label{eq:eom_int}
    \frac{1}{2N}\diby{}{\lambda}\lf( \frac{1}{2N} g_{ab}(\bar{q})\, \dot{\bar{q}}^b \rt) = -\frac{1}{2} \bar{p}_b \bar{p}_c \bar{\partial}_a g^{bc}(\bar{q}).
}

We can now use the conformal flatness of the metric $g_{ab} = e^\phi \eta_{ab} = (-2V) \eta_{ab}$ and the scalar constraint $g^{ab}\, p_a p_b = 1 \ra \eta^{ab} p_a p_b = -2V$ to write (\ref{eq:eom_int}) in a more recognizable form. Identifying $\dot\tau \equiv -\frac{N}{V}$, (\ref{eq:eom_int}) reduces to
\equa{\label{eq:newton}
    \frac{\partial^2 \bar{q}^a}{\partial\tau^2} = -\bar{\partial}^a V(\bar{q}).
}
This is Newton's $2^\text{nd}$ law with $\tau$ playing the role of Newtonian time and with the $q$'s replaced by $\bar{q}$'s. Note that we did not use the conformal flatness of the metric until the last step and then only to write our results in a more recognizable form. We note in passing that Newton's laws are just \eq{newton} written in the \emph{proper time} gauge, analogous to the similar gauge condition used in general relativity, where $N=1$ and $\Na = 0$. This special gauge also corresponds to Barbour's \emph{distinguished representation} \cite{barbour:scale_inv_particles}.

\subsubsection{Solving the Constraints}

It is now possible to use Hamilton's first equation to invert the scalar and vector constraints and solve explicitly for the lapse and shift. This will allow us to write the equations of motion in terms of gauge invariant quantities having eliminated all auxiliary fields $\omega$. Solving for the lapse and shift tells what gauge should be used in order to satisfy both the equations of motion and the initial and final conditions imposed on the $q$'s.

The shift can be solved for by inserting Hamilton's first equation
\equa{
    \dot q^a = 2N p_b g^{ab} - \Na \lf. t_\alpha \rt.^a_b q^b
}
into the vector constraint $\mathcal H_\alpha = \pia - p_a \ta{a}{b} q^b = 0$ after applying the Mach condition $\pia = 0$. Inverting the result for $\Na$ gives
\equa{\label{eq:shift}
    \Na M_{\alpha\beta} = \eta_{ab} \dot{q}^a \lf. t_\beta \rt.^b_c q^c,
}
where
\equa{
    M_{\alpha\beta} = \eta_{ab} \ta{a}{c} \lf. t_\beta \rt.^b_d\, q^c q^d.
}
In the above, we have used $g_{ab} = e^\phi \eta_{ab}$ and removed as factors $e^\phi$ and $N$. The fact that $N$ drops out is what allows the scalar and vector constraints to decouple allowing the system to be easily solved. The field theories are typically more sophisticated, and this is no longer possible. Being symmetric, $M_{\alpha\beta}$ is invertible. Thus, $\Na$ is given formally using the inverse $M^{\alpha\beta}$ of $M_{\alpha\beta}$. In \scn{bid_observables} we shall give simple closed--form expressions for $\Na$ for non--relativistic particle models invariant under translations and dilitations. The inversion of $M_{\alpha\beta}$ for non--Abelian groups, such as the rotations in 3 dimensions, is formally possible but illuminating, closed--form expressions are difficult to produce.

The lapse can be solved for using (\ref{eq:killing_field}) and inserting Hamilton's first equation (\ref{eq:h1_pbar}) into the scalar constraint $\mathcal H = g^{ab} p_a p_b -1 = 0$. This gives
\equa{\label{eq:lapse}
    N = \frac{1}{2} \sqrt{g_{ab}(\bar{q})\, \dot{\bar{q}}^a\dot{\bar{q}}^b}.
}
Having already solved for the shift we can use it to compute $G^a_b(\wa)$ in the above expression using the equation of motion $\dw = \Na$. We can now express all equations of motion without reference to auxiliary quantities.


\subsection{Gauge--Independent Observables} \label{sec:bid_observables}

The simple form of \eq{eom_int} and (\ref{eq:newton}) suggests there might be something fundamental about the corrected coordinates $\bar{q}^a = G^a_b q^b$. In fact, as can be easily checked, they commute with the primary, first class vector constraints $\lin$. The $\bar{q}$'s are then invariant under the gauge transformations \eq{gauge_trans} generated by $\lin$. They do \emph{not}, however, commute with the quadratic scalar constraint. For this reason, they are non--perennial \emph{observables} in the language of Kucha\v r \cite{kuchar:time_int_qu_gr,Kuchar:can_qu_gra} who argues that such quantities are \emph{the} physically meaningful observables of reparameterization invariant theories. Barbour and Foster take this argument further in \cite{barbour_foster:dirac_thm} showing how Dirac's theorem fails for finite--dimensional reparameterization invariant theories. The reason for ignoring the non--commutativity of the observables with the scalar constraint is that the scalar constraint generates physically \emph{distinguishable} configurations. This is in contrast to the linear vector constraints, which generate physically \emph{indistinguishable} states. In this work, we will use Kucha\v r's language to describe these observables and see that, in all cases where the constraints can be solved explicitly, the $\bar{q}^a$ are manifestly relational observables.

The form of $\bar{q}^a$ and $\lin$ is critical for their commutativity: the $q^b$ part of $\bar{q}^a$ fails to commute with the $p^a\ta{a}{b}q^b$ piece of $\lin$ by exactly the amount required to cancel the non--commutativity of the $G^a_b$ piece with $\pia$. If we were to incorrectly apply the Mach condition $\pia = 0$ \emph{before} computing the PBs, we would obtain the false conclusion that the $\bar{q}^a$ are not observables in the sense defined above. This highlights an important advantage of treating the auxiliary fields $\wa$ as \emph{cyclic} variables\footnote{This situation is slightly more complicated when the action is not initially globally gauge invariant.} with a Mach variation rather than treating them as Lagrange multipliers, as is done, for example, in the ADM theory. Treating the $\omega$'s as Lagrange multipliers produces equivalent classical equations of motion but hides the fact that the $\bar{q}^a$'s are genuine observables. Thus, best matching establishes what are the true degrees of freedom.

The corrected coordinates, $\bar{q}^a$, have a nice geometric interpretation. Using Hamilton's first equation for $\dw$, the corrected coordinates can be written in terms of the shift as
\equa{\label{eq:holonomy}
    \bar{q}^a = \myexp{\wa\ta{a}{b}}q^b = \mathcal{P}\myexp{\int \Na\ta{a}{b}\, d\lambda} q^b,
}
where $\mathcal P$ implies path--ordered integration. Thus, the corrected coordinates are obtained by subtracting the action of the open--path holonomy of the lapse (thought of as the pullback of the connection over $\acs$ onto the classical path) on the $q^b$'s. This subtracts all vertical motion of the $q$'s along the fiber bundle.

\subsubsection{Special Cases}

The significance of the $\bar{q}$'s is more clearly seen by solving the constraints for specific symmetry groups. First, consider non--relativistic particle models invariant under translations. The $q$'s represent particle positions in 3 dimensional space. The $a$ indices can be split into a vector index, $i$, ranging from 1 to 3, and a particle index, $I$, ranging from 1 to the total number of particles in the system. Then, $a = iI$. For $\eta_{ab}$ we use the diagonal mass matrix attributing a mass $m_I$ to each particle. With the generators of translations, \eq{gen_trans}, (\ref{eq:shift}) takes the form
\equa{
    \vec{N} = \frac{\sum_I m_I \vec{q}_I}{\sum_I m_I} \equiv \dot{\vec{q}}_{\text{cm}}.
}
The shift is the velocity of the center of mass $\vec{q}_{\text{cm}}$. Aside from an irrelevant integration constant, which can be taken to be zero, the auxiliary fields $\omega^\alpha$ represent the position of the center of mass. Inserting this result into \eq{holonomy}, the corrected coordinates are
\equa{
    \bar{q}^a = q^a - q^a_{\text{cm}}.
}
They represent the difference between the particles' positions and the center of mass of the system. This is clearly a relational observable. Furthermore, the non--physical quantity is the position of the center of mass since the theory is independent of its motion.

We can also treat models invariant under dilatations.\footnote{See \cite{barbour:scale_inv_particles} for more details on these models.} In this case, (\ref{eq:shift}) is easily invertible since there is only a single shift function, which we will call $s$. Using the same index conventions as before and the generators \eq{gen_dil} we find
\equa{
    s = \diby{}{\lambda} \lf( -\frac{1}{2}\ln I \rt),
}
where $I = \sum_I m_I (\vec{q}_I)^2$ is the moment of inertia of the system. Aside from an overall integration constant, which can be set to zero, the auxiliary field is $-1/2$ times the log of the moment of inertia. Using \eq{holonomy}, the corrected coordinates are the original coordinates normalized by the square root of the moment of inertia
\equa{
    \bar{q}^a = \frac{q^a}{\sqrt{I}}.
}
Because $I$ contains two factors of $q$, $\bar{q}$ will be invariant under rescalings of the coordinates. Thus, the corrected coordinates are independent of an absolute scale.

The quantity $\tau$, which plays the role of the Newtonian time, can now be computed. It is a function of the lapse and the potential. Since the lapse is an explicit function of the corrected coordinates, it will be observable. Using the definition $\dot\tau \equiv -\frac{N}{V}$ and (\ref{eq:lapse}), $\tau$ is simply
\equa{
    \tau = \int d\lambda\, \sqrt{-\frac{T(\bar{q})}{V(\bar{q})}},
}
where $T = \frac{1}{2} m_{ab} \dot{\bar{q}}^a\dot{\bar{q}}^b$ is the relational kinetic energy of the system. $\tau$ is independent of $\lambda$ and observable within the system. Thus, once the constraints have been solved for, the equations of motion (\ref{eq:newton}) are in a particularly convenient gauge--independent form. This definition of $\tau$ corresponds to BB's \emph{ephemeris time} \cite{barbour:nature, barbourbertotti:mach}.

\subsection{Background Dependence and Independence}\label{sec:BI_BD}

The presence of a symmetry of the configurations of $\acs$ allows for a distinction between two types of theories:\footnote{When there is a symmetry of the configurations which is not reflected in a global symmetry of the metric on $\acs$, one can \emph{construct} a gauge invariant metric using additional auxiliary fields. This situation, which is considered in \cite{barbour_el_al:physical_dof,barbour_el_al:scale_inv_gravity}, is more complicated than that of this paper. Nevertheless, I believe these definitions are sufficiently general to hold in this case as well.}
\begin{itemize}
  \item those that attribute physical significance to the exact location of the configuration variables along the fiber generated by the symmetry. We will call these theories \emph{Background Dependent (BD)}.
  \item those that \emph{do not} attribute any physical significance to the exact location of the configuration variables along the fiber. These theories will be called \emph{Background Independent (BI)}.
\end{itemize}
Based on these definitions, it would seem odd even to consider BD theories as they distinguish between members of an equivalence class. These theories are useful nevertheless whenever there is an \emph{emergent} background that breaks the symmetry in question at an effective level.

For a historically relevant example of why BD theories are important, consider Newton's well known bucket argument. In his \emph{Principia}, Newton argues that the relative motion between a spinning bucket and the water that it holds cannot explain the precise way in which the water creeps up the walls of the bucket. He concludes that only the bucket and water's motion through \emph{absolute} space can explain the behavior. This serves as justification for the background dependence of Newton's theory with respect to rotations. No one would argue that Newton's mechanics is not useful, at least at an effective level. Nevertheless, in a modernized version of Mach's well known rebuttal, one could argue that this BD theory should be emergent out of a fundamentally BI theory that takes into account the relative motion of the water and bucket with the rest of the universe. From the perspective of best matching, since the moment of inertia of the water--bucket system is so small compared to that of the fixed stars, the water and bucket could have any reasonable value of angular momentum without disturbing the total angular momentum of the universe, which, as we have seen, is constrained to be zero. This example illustrates why, although it makes sense to treat only BI theories as fundamental, BD theories are, nevertheless, very \emph{useful} in practice.


Best matching provides a framework for making our definitions of BD and BI more precise. Whenever there are symmetries in the configurations  it is possible to introduce auxiliary fields $\wa$ whose role is nothing more than to parametrize the symmetry. Indeed, making the $\omega$'s dynamical can be very useful since, as we have seen, the full power of Dirac's formalism \cite{dirac:lectures} can be used to study the dynamical effects of the symmetry. In addition, introducing the $\wa$ fields gives us the freedom to distinguish between BD and BI theories as follows:
\begin{itemize}
  \item \emph{BD theories} are those that vary the $\wa$ fields in the standard way using \emph{fixed} endpoints. This requires the specification of appropriate initial and final data, which is considered to be physically meaningful.\footnote{So as not to add redundancy to the boundary conditions, we can set $\wa(\lambda_{\text{in}}) = \wa(\lambda_{\text{fin}}) = 0$ without loss of generality. The boundary conditions on the $q$'s will then contain all the information about the absolute position of the $q$'s along the fiber.}
  \item \emph{BI theories} are those that vary the $\wa$ fields using a Mach variation.
\end{itemize}

Given these definitions, we can understand the physical difference between BD and BI theories by considering the form of the vector constraints (\ref{eq:lin}). In the BD case, the $\pia$'s are constants of motion. These constants are determined uniquely by the initial conditions on the $q$'s. However, in the BI theory, the constants of motion are irrelevant and are seen as unphysical. The initial data cannot affect their value. As a result, the BD theory requires more inputs in order to give a well defined evolution. The difference is given exactly by the dimension of the symmetry group. This is precisely in accordance with Poincar\' e's principle.

\subsection{Time and Parametrized Hamilton's Principle}

Parametrized Hamilton's Principle (PHP) is an alternative to Jacobi's principle for determining the dynamics of a system. It is still a geodesic principle on configuration space but the square root is disposed of in place of a mathematically simpler action. The cost of having this simpler action is the introduction of an auxiliary field whose role is to restore the reparameterization invariance. PHP has the advantage over Jacobi's principle that it singles out a preferred parametrization of the geodesics through a choice of normalization of the scalar constraint. For a standard normalization, the preferred parameter is just BB's ephemeris time. In geometrodynamics, this quantity will be related to the proper time of a freely falling observer. In the case of the particle models, PHP is just the standard model of parameterized particle dynamics treated, for example, in \cite{lanczos:mechanics}.

Because time is now dynamical, we can use the definitions of BI and BD from \scn{BI_BD} to distinguish between theories that have a background time and those that are timeless. As we would expect, Newton's theory, which contains explicitly an absolute time, can be obtained from PHP with a background time. Alternatively, Jacobi's timeless theory is obtained from PHP by keeping the time background independent.

\subsubsection{Action and Hamiltonian}

To simplify the discussion, we will ignore for the moment the spatial symmetries. This will avoid having to deal with the linear constraints. Comparison to the equations of previous sections can either be made by setting the shift, $\Na$, equal to zero or by unbarring quantities. It can be verified that neglecting the spatial symmetries does not affect the discussions regarding time  \cite{sg:mach_time}.

PHP is defined by the action
\equa{\label{eq:ham_action}
	S_H = \int_{q_{\text{in}}}^{q_{\text{fin}}} d\lambda\, \frac{1}{2}\lf[ \frac{1}{\dot\tau} \gamma_{ab} \dot{q}^a \dot{q}^b + \dot\tau e^\phi \rt].
}
The Lagrangian has the form of Hamilton's principle $T - V$ (recall that $e^\phi = -2V$) but the absolute time $\tau$ has been promoted to a dynamical variable by parameterizing it with the auxiliary variable $\lambda$. This explains the name: Parametrized Hamilton's Principle. The particular normalization used takes advange of the conformal split of the metric and singles out BB's ephemeris time as a preferred parameter for the geodesics.\footnote{Alternatively, one could split the action as $S_H = \int d\lambda\, \lf[ \frac{1}{\dot\tau} e^\phi\gamma_{ab} \dot{q}^a \dot{q}^b - \dot\tau \rt]$ without changing the equations of motion. This action would single out an affine parameter for the geodesics. It corresponds to multiplying the scalar constraint by $e^{-\phi}$.} For simplicity, we restrict ourselves to metrics of the form: $\gamma_{ab} = \eta_{ab}$.

We can perform a Legendre transform to find the Hamiltonian of the system. Defining the momenta
\begin{align}
	p_a &= \frac{\delta S_H}{\delta \dot{q}^a} = \frac{1}{\dot\tau} \eta_{ab} \dot q^b, \qand \label{eq:p_a}\\
	p_0 &= \frac{\delta S_H}{\delta \dot\tau} = -\frac{1}{2} \lf[ \frac{\eta_{ab}\dq^a\dq^b}{\dot\tau^2} - e^\phi \rt] \label{eq:p_0}
\end{align}
we note that they obey the scalar constraint
\equa{
	\ham_{\text{Ham}} \equiv \frac{1}{2}\lf( \eta^{ab} p_a p_b - e^\phi \rt) + p_0 = \frac{e^\phi}{2} \ham_\text{Jacobi} + p_0 = 0.
}
The appearance of the $p_0$ term is the only difference, other than the factor $e^\phi$, between Jacobi's principle and PHP (the factor 2 is purely conventional). The factor $e^\phi$ can be absorbed by a field redefinition of the lapse and has no bearing on physical observables. It is a relic of our choice of ephemeris time to parametrize geodesics. Using the definitions \eq{p_a} and \eq{p_0}, we find that the canonical Hamiltonian is identically zero, as it must be for a reparameterization invariant theory. Thus, the total Hamiltonian is
\equa{\label{eq:pnm_ham}
	H_\text{T} = N\ham = N \lf( \frac{1}{2}\eta^{ab} p_a p_b -\frac{1}{2} e^\phi + p_0 \rt).
}

\subsubsection{BI Theory}

In PHP, time is promoted to a configuration space variable. The symmetry associated with translating the origin of time is reflected in the invariance of the action under time translations $\tau \ra \tau + a$, where $a$ is a constant. In \cite{sg:mach_time}, it is shown that applying the best--matching procedure to this symmetry is equivalent to treating $\tau$ itself as an auxiliary field. To make the theory background independent with respect to the temporal symmetries, we follow the procedure outlined in \scn{BI_BD} and impose the Mach condition after evaluating the Poisson brackets. In this case, the Mach condition takes the form $p_0 = 0$.

We pause for a brief observation. Since the Mach variation of a cyclic variable is equivalent to the standard variation of a Lagrange multiplier, we can replace $\dot\tau$ with $N$ when doing a background independent formulation of PHP. Then, the action (\ref{eq:ham_action}) bears a striking resemblance to the ADM action. This illustrates why the ADM action is background independent as far as time is concerned. However, the ADM action hides the possibility of introducing a background time (following the procedure given in the next section). Considering this, it might be more enlightening to think of the lapse as a cyclic variable subject to Mach variation as is done in \cite{barbour:scale_inv_particles} and \cite{Anderson:cyclic_ADM}.

In order to compare this to the Jacobi theory, it is instructive to work out the classical equations of motion
\begin{align}
 	\dq^a &= \pb{q^a}{H_\text{T}} = N \eta^{ab} p_b, \\
	\dot p_a&= \pb{p_a}{H_\text{T}} = N \partial_a \lf( \frac{e^\phi}{2} \rt) = -N \partial_a V  \\
	\dot\tau &= \pb{\tau}{H_\text{T}} = N, \qand \label{eq:ham1_tau}\\
	\dot p_0 &= \pb{p_0}{H_\text{T}} = 0. \label{eq:ham_p0}
\end{align}
(\ref{eq:ham1_tau}) reinforces the fact that $\tau$ is an auxiliary. It is straightforward to show that the above system of equations implies
\equa{
	\frac{\partial^2 q^a}{\partial\tau^2} = -\partial^a V(q).
}
This is Newton's $2^{\text{nd}}$ law. Solving the scalar constraint gives an explicit equation for $\tau$,
\equa{\label{eq:ham_tau}
	\dot\tau = \sqrt{\eta_{ab} \dq^a \dq^b e^{-\phi}} = \sqrt{\frac{T}{-V}},
}
using the definitions for $V$ and $T$ given in \scn{jacobi_principle}. This is precisely the expression for the ephemeris time $\tau$ defined in the Jacobi theory. It should be noted that the Mach condition implies that the integration constant of (\ref{eq:ham_p0}) is zero. Use was made of this to deduce (\ref{eq:ham_tau}). From this it is clear that the BI theory is classically equivalent to Jacobi's theory.

One can take this further and compare the two theories quantum mechanically. Noticing that the canonical action is linear in $\dot\tau$ and $p_0$, we can integrate out $\tau$ without affecting the quantum theory and use the Mach condition $p_0 = 0$ to reduce the scalar constraint to
\equa{
	\ham = g^{ab} p_a p_b - 1 = 0
}
(after factoring $e^\phi$). This is the scalar constraint (\ref{eq:jacobi_constraint}) of Jacobi theory. With $\tau$ now defined by (\ref{eq:ham_tau}), the canonical theories are identical. Thus, their canonical quantizations should also match. For more details on the equivalence of these theories quantum mechanically, see \cite{sg:mach_time}, where the path integrals for these theories are worked out in detail.

\subsubsection{BD Theory}

For the BD theory, we do not impose the Mach condition. Integration of (\ref{eq:ham_p0}) implies $p_0 \equiv -E$. The only effect that this has on the classical theory is to alter the formula for $\tau$ to
\equa{
	\dot\tau = \sqrt{\frac{T}{E-V}}.
}
Now an initial condition is imposed on $\tau$ that fixes the value of $E$ and violates Poincar\' e's principle. Thus, $\tau$ is equivalent to a Newtonian absolute time. Note that inserting a background time would have been impossible if we started with the ADM form of PHP.

Strictly speaking, there is a difference between $E$, defined as the negative of the momentum canonically conjugate to time, and $E'$, which is just the constant part of $V = -E' + V'$. Together they form what we would normally think of as the total energy $E_\text{tot} = E + E'$ of the system. $E'$ is freely specifiable and plays the role of a fundamental constant of nature while $E_\text{tot}$ is fixed by the initial conditions on $\tau$. In the classical theory, it is unnecessary to make a distinction between $E_\text{tot}$ and $E$. However, in the quantum theory, this distinction is important because of the possible running of constants of nature like $E'$. In general relativity, the role of $E'$ is played by the cosmological constant. As a result, this distinction may be relevant to the cosmological constant problem \cite{Smolin:unimodular_grav}.

\subsubsection{A Problem of Time}

In the classical theory, it seems that there is only a very subtle difference between the BI and BD theories. The difference amounts to the ability to impose boundary conditions on $\tau$ that constrain the total energy. However, the quantum theories are drastically different. Using Dirac's procedure, we promote the scalar constraint to an operator constraint on the wavefunction $\Psi$. In the BD theory, Dirac's procedure applied to the Hamiltonian (\ref{eq:pnm_ham}) gives the time dependent Schr\"odinger equation
\equa{
	\hat\ham \Psi = \lf[ \frac{1}{2}\eta^{ab} \hat p_a \hat p_b + V(\hat q) + \hat p_0 \rt] \Psi  = 0.
}
In a configuration basis, $p_0 = -i\diby{}{\tau}$. Thus, the above is indeed the standard Schr\"odinger equation. However, in the BD theory, the Mach condition requires $p_0 = 0$ leaving instead the time \emph{independent} Schr\"odinger equation
\equa{
	\hat\ham \Psi = \lf[ \frac{1}{2}\eta^{ab} \hat p_a \hat p_b + V(\hat q) - E' \rt] \Psi  = 0,
}
where we have explicitly removed the constant part of the potential. While it is easy to define an inner product in the BD theory under which evolution will be unitary this is not the case in the BI theory. This makes it difficult to define a Hilbert space for the BI theory (at least at the level of the entire universe). The difficulties associated with this can be called a \emph{problem of time} similar to what happens in quantum geometrodynamics.\footnote{In geometrodynamics, there are \emph{additional} complications associated with foliation invariance or many-fingered time. As far as I know, these have no analogues in the finite--dimensional models.} It is interesting to note that, in finite--dimensional models, one can eliminate this problem of time by artificially introducing a background time. In \scn{unimodular}, we study the effects of applying the same procedure to geometrodynamics and are led to unimodular gravity. Clearly the issue of background independence is of vital importance in the quantum theory. This will have important implications in any quantum theory of gravity.

\section{Infinite Dimensional Relational Models}

We will now consider field theories over a spatial manifold $\Sigma$. It will be sufficient for $\Sigma$ to be an $n$--dimensional manifold with Euclidean signature. For simplicity, we will assume that $\Sigma$ is closed with no boundary. The spatial dependence of the configurations leads to an ambiguity in how to take the square root in Jacobi's principle. There are two choices: 1) integrate over space first then take the square root at every $\lambda$ or 2) take the square root first then integrate over space and repeat this for every $\lambda$. The first option is highly non--local and leads to theories with a preferred time foliation. It is particularly adapted to theories with a \emph{projectable} lapse such as the theory proposed by Ho\v rava \cite{Horava:gravity1,Horava:gravity2}. The second option is a local action principle and leads to local theories such as general relativity. We will briefly consider the global square root theories below then treat the local theories in more detail.

\subsection{Global Square Root Theories}\label{sec:global_sqrt}

\subsubsection{Global Jacobi Action}

Instead of considering the most general case of arbitrary fields defined over a manifold $\Sigma$, we will treat the specific case of dynamical geometries. The configurations are the symmetric 2--forms, $g_{ab}$, defining a metric on $\Sigma$. This allows us to consider a general class of geometrodynamic theories. The configuration space is Riem: the space of all possible metrics on $\Sigma$.

Metrics \emph{on Riem} should be functionals of the spatial metric, $g$, and should feed on two symmetric 2--forms, $u$ and $v$. For an up to date account of how to define metrics on Riem, see \cite{Giulini:superspace}. We will only consider those metrics $\mathcal G$ that split into an ultra--local piece
\equa{\label{eq:gen_dewitt}
	G[u, v, g] \equiv \int_\Sigma d^nx \sqrt{g}\, G^{abcd}(x)u_{ab}(x) v_{ab}(x) \equiv \int_\Sigma d^nx \sqrt{g}\, (g^{ac} g^{bd} - \alpha g^{ab}g^{cd})u_{ab} v_{cd},
}
and a conformal piece $\mathcal V[g, \nabla g, \hdots] = \int d^nx\,\sqrt{g} V$ such that $\mathcal{G}[u,v, g,\nabla g,\hdots] = \mathcal V[g, \nabla g, \hdots]\cdot G[u, v, g]$. Note that $G^{abcd}$ is the most general ultra--local rank--4 tensor that can be formed from the metric. It represents a one parameter family of supermetrics labeled by $\alpha$. For $\alpha = 1$, we recover the usual DeWitt supermetric. $G^{abcd}$ plays a similar role to the flat metric $\eta_{ab}$ in the finite--dimensional theories. The scalar function $V(g(x),\nabla g(x),\hdots)$ is analogous to the conformal factor of the finite--dimensional theories and, for this reason, is often called the potential. However, it differs from the potential of the finite--dimensional models in that it can depend on the \emph{spatial} derivatives of the metric. 

The symmetry of the configurations is with respect to spatial diffeomorphisms. This can be reflected in the action by requiring it be a spatial scalar. We can best match this symmetry by introducing the corrected coordinates
\equa{
	\bar{g}_{ab} = \myexp{\mathcal{L}_\xi} g_{ab}
}
and doing a Mach variation with respect to the auxiliary fields $\xi$. As was shown in general for the finite--dimensional models, this is equivalent to introducing the gauge covariant derivative
\equa{
	\mathcal{D}_\xi g_{ab} = \dot{g}_{ab} + \mathcal{L}_{\dot\xi} g_{ab} = \dot{g}_{ab} + \dot\xi_{(a;,b)},
}
which replaces all occurrences of $\frac{d}{d\lambda}$ in the action. In the above, semi--colons represent covariant differentiation on the tangent bundle of $\Sigma$ using a metric compatible connection.

We can now write down a Jacobi--type action for this theory. A direct analogy with the finite--dimensional models gives
\begin{align}
  S_{\text{global}} &= \int d\lambda \sqrt{\mathcal G[\mathcal{D}_\xi g,\mathcal{D}_\xi g, g, \nabla g, \hdots]} \notag \\
		    &= \int d\lambda \sqrt{\int_\Sigma d^n x \sqrt{g}\, G^{abcd} \mathcal{D}_\xi g_{ab} \mathcal{D}_\xi g_{cd}} \cdotp\sqrt{\int_\Sigma d^n x' \sqrt{g}\, V(g,\nabla g, \hdots)}.\label{eq:global_jacobi}
\end{align}
Clearly, \eq{global_jacobi} is a non--local action as it couples all points in $\Sigma$ at a given instant.

\subsubsection{Projectable--Lapse Theories}

The momenta obtained from the action (\ref{eq:global_jacobi}) are
\begin{align}
    \pi^{ab} &= \frac{\delta S}{\delta \dot g_{ab}} = \sqrt{\frac{\mathcal V}{\mathcal T}} \sqrt{g} G^{abcd} \dxi g_{cd} \\
    \zeta^{a} &= \frac{\delta S}{\delta \dot \xi_{a}} = -2\nabla_b \lf( \sqrt{\frac{\mathcal V}{\mathcal T}} \sqrt{g}G^{abcd} \dxi g_{cd} \rt),
\end{align}
where $ \mathcal T = G[\dxi g,\dxi g, g] = \int d^nx\,\sqrt{g} G^{abcd} \dxi g_{ab}\, \dxi g_{cd}$ is the kinetic term. This leads to the primary constraint
\equa{
    \mathcal H^a(x) = \zeta^a(x) + 2\nabla_b \pi^{ab}(x) = 0,
}
which, combined with the Mach condition $\zeta^a = 0$, is just the standard diffeomorphism constraint of general relativity. Although this constraint is clearly local, there is a second primary constraint that is only true when integrated over all of space. This constraint is the \emph{zero mode} of the usual Hamiltonian constraint
\equa{
    \mathcal H^{(0)} = \int d^nx\, \lf[ \frac{1}{\sqrt g} G_{abcd} \pi^{ab} \pi^{cd} - \sqrt{g} V \rt] \equiv \int d^nx\,\mathcal H.
}
It guarantees that the metric on Riem $\mathcal G[\dxi g,\dxi g, g, \nabla g, \hdots]$ is non--negative.

The total Hamiltonian is
\equa{
    H_\text{tot} = N(\lambda) \mathcal H^{(0)} + \int d^nx N^a(\lambda,x) \mathcal H_a.
}
The lapse function is only $\lambda$, and not $x$, dependent. It is said to be projectable. Because of this, the theory does not obey the full Dirac--Teitelboim algebra \cite{Teitelboim:DT_algebra} and is invariant only under foliation preserving diffeomorphism and not the full $n+1$ diffeomorphism group. Despite this, these theories can still be very useful and, depending on the choice of potential, can represent either symmetry reduced versions of general relativity or Lorentz invariance violating theories like those considered by Ho\v rava \cite{Horava:gravity1,Horava:gravity2}.

\subsection{Local Square Root Theories}

In this section we explore theories that take the square root before integrating over all of space. Physically, this seems like the more natural choice because the action principle is now local. On the other hand, the mathematical structure is less appealing because we no longer have a proper metric on Riem and we loose a direct analogy with the finite--dimensional models. We can no longer write the action in terms of a quantity that gives the ``distance'' between two infinitesimally separated geometries. Furthermore, using a local square root produces a local scalar constraint that restricts one degree of freedom at every point. With the right choice of potential, this extra gauge freedom manifests itself as foliation invariance and leads to many technical and conceptual issues, particularly in the quantization. For a review of the difficulties associated with foliation invariance and other issues associated to time, see \cite{kuchar:time_int_qu_gr,Isham:pot_review}. Despite these complications, examples of local square root theories include general relativity. Thus, it seems Nature has forced them upon us.\footnote{Provided GR is the correct theory of spacetime at all energies.}

\subsubsection{Geometrodynamics from Jacobi's Principle}\label{sec:geo_jacobi}

Bringing the square root inside the spatial integration while keeping a structure analogous to Jacobi's principle for the finite--dimensional models gives
\equa{\label{eq:local_geo}
  S_{\text{local}} = \int d\lambda\, d^nx \, \sqrt{g\,G^{abcd} \mathcal{D}_\xi g_{ab} \mathcal{D}_\xi g_{cd}\cdotp V(g,\nabla g)},
    }
where $G^{abcd} = g^{ac}g^{bd} - \alpha g^{ab}g^{cd}$. The quantity $V\cdot G^{abcd}$ is the infinitesimal ``distance'' between two \emph{points} of two infinitesimally separated geometries. It is a kind of pointwise metric on Riem. Thus, there is no clean geodesic principle on the reduced configuration space in contrast to either the finite--dimensional case or the global square root theories.

The action \eq{local_geo} has been analysed in detail in \cite{barbour_el_al:physical_dof,barbourbertotti:mach,barbour_el_al:scale_inv_gravity}. For the special choices $\alpha = 1$ and
\equa{
  V(g,\nabla g,\nabla^2 g) = 2\Lambda - R(g,\nabla g,\nabla^2 g),
}
where $R$ is the scalar curvature of $\Sigma$ and $\Lambda$ is a constant, the constraint algebra is known to close. With these choices, \eq{local_geo} is the Baierlein--Sharp--Wheeler (BSW) action of GR \cite{bsw:bsw_action}, whose Hamiltonian equations of motion are equivalent to those of ADM \cite{adm:adm_review}. Thus, Jacobi's principle with local square root naturally recovers GR.\footnote{Equivalence is achieved because the Mach variation of $\dot\xi$ is equivalent to the usual variation of the shift vector $N^i$. See, for example, \cite{Anderson:cyclic_ADM}.}



\subsubsection{Geometrodynamics with Parametrized Hamilton's Principle} \label{sec:unimodular}

While Jacobi's principle combined with the best matching of the $n$-diffeomorphism invariance leads to the BSW action of GR, it would be nice if there was a natural way to obtain the usual ADM formulation of GR from relational principles. This is provided by PHP (for a demonstration of this following a Ruthian reduction see \cite{Anderson:cyclic_ADM}). Furthermore, since it explicitly includes an auxiliary time and singles out the ephemeris time as a preferred geodesic parameter, PHP provides a natural framework for introducing a notion of background time in GR. Interestingly, this procedure leads directly to unimodular gravity.

To implement PHP, we use the kinetic term and potential outlined in \scn{geo_jacobi}. Using a local action principle and introducing the auxiliary field $\tau^0(\lambda, x)$, the analogue of (\ref{eq:ham_action}) is
\equa{\label{eq:geo_ham}
	S_H = \int d\lambda\, d^n x\, \sqrt{g} \frac{1}{2} \lf[ \frac{1}{\dot\tau^0} G^{abcd} \mathcal{D}_\xi g_{ab} \mathcal{D}_\xi g_{cd} - \dot\tau^0 (2\Lambda' - R) \rt],
}
where we have used a prime to distinguish $\Lambda'$ from another $\Lambda$ that we will consider later. This is completely analogous to $E$ versus $E'$ encountered in the particle models. It can be verified that using a local function, $\tau^0$, of $x$ is equivalent to taking a local square root in the Jacobi action. Similarly, a global $\tau^0$ is equivalent to a global square root. 


The ADM theory can be obtained by doing a short canonical analysis of the action \eq{geo_ham}. The momenta are:
\begin{align}
	\pi^{ab} &= \diby{L}{\dot{g}_{ab}} = \frac{\sqrt{g}}{\dot\tau^0} G^{abcd}(\dot{g}_{cd} + \dot\xi_{(c,d)}), \\
	\zeta^a &= \diby{L}{\dot\xi_a} = -\nabla_b \lf( \frac{\sqrt{g}}{\dot\tau^0} G^{(ab)cd}(\dot{g}_{cd} + \dot\xi_{(c,d)}) \rt), \qand \\
	p_0 &= \diby{L}{\dot\tau^0} = -\frac{\sqrt{g}}{2}\lf( \frac{1}{(\dot\tau^0)^2} G^{abcd} \mathcal{D}_\xi g_{ab} \mathcal{D}_\xi g_{cd} + (2\Lambda' -R) \rt).
\end{align}
The scalar constraint is
\equa{\label{eq:ham_constraint_ham}
	\ham = \frac{1}{\sqrt{g}} G_{abcd}\pi^{ab}\pi^{cd} + \sqrt{g}(2\Lambda'-R) +  2p_0 = \hamadm + 2p_0 = 0,
}
where $\hamadm$ is just the scalar constraint of the ADM theory. There is also a vector constraint associated with $\zeta^a$. It is
\begin{equation}
 	\diff = \nabla_b \pi^{(ab)} + \zeta^a = \diffadm + \zeta^a = 0.
\end{equation}
$\diffadm$ is ADM's usual vector constraint.

The canonical Hamiltonian is zero as it should be in a reparameterization invariant theory. Thus, the Hamiltonian is
\equa{\label{eq:ham_tot_ham}
	H = N\ham + N_a \diff = H_\text{ADM} + 2Np_0 + N_a \zeta^a.
}
$H_\text{ADM}$ is the ADM Hamiltonian. However, this may not be the full Hamiltonian since we need to check for secondary constraints. To do this, we introduce the fundamental equal--$\lambda$ PB's
\begin{align}
  \pb{g_{ab}(\lambda,x)}{\pi^{cd}(\lambda, y)} &= \delta^c_a\delta^d_b\,\delta(x,y), \\
  \pb{\xi_a((\lambda,x)}{\zeta^b(\lambda, y)} &= \delta^b_a\,\delta(x,y), \qand \\
  \pb{\tau^0(\lambda,x)}{p_0(\lambda,y)} &= \delta(x,y).
\end{align}
Then, the constraint algebra reduces to
\begin{align}
  \pb{g^{-1/2} \ham(x)}{\ham(y)} &= \lf[ (g^{-1/2}\diffadm)(x) + (g^{-1/2}\diffadm)(y) \rt] \delta(x,y)_{;a} \\
  \pb{g^{-1/2} \ham(x)}{\diffadm(y)} &= g^{-1/2} \hamadm(x)^{;a} \delta(x,y) \label{eq:ham_diff}\\
  \pb{g^{-1/2} \diff(x)}{\ham^b(y)} &= (g^{-1/2}\diffadm)(x)\delta(x,y)^{;b} + (g^{-1/2}\hamadm^b)(y)\delta(x,y)^{;a}.
\end{align}
At this point, the discussions for standard and Mach variations diverge.

\subsubsection{Mach Variation: Time--Independent Theory}

After taking PB's we can apply the Mach conditions for $p_0$ and $\zeta^a$
\begin{align}
  p_0\approx & 0 \\
  \zeta^a\approx & 0.
\end{align}
Then, the vector and scalar constraints imply
\begin{align}
  \hamadm\approx & 0 \\
  \diffadm\approx & 0.
\end{align}
Thus, the constraint algebra is first class and the total Hamiltonian is given by \eq{ham_tot_ham}.

At this point, we can't use the Mach conditions to recover the ADM theory because they are only weak equations. To see that the ADM theory is indeed recovered, we work out the classical equations of motion. The terms in \eq{ham_tot_ham} that are new compared with the ADM theory commute with $g_{ab}$ and $\pi^{ab}$. Thus, they do not affect the equations of motion for $g_{ab}$ or for $\pi^{ab}$ other than replacing the lapse $N$ with $\dot\tau^0$ and the shift $N_a$ with $\dot\xi_a$. Since the remaining equations of motion just identify
\begin{align}
 	\dot\tau^0 &= \pb{\tau^0}{\htot} = 2N, \qand \\
	\dot\xi_a & = \pb{\xi_a}{\htot} = N_a,
\end{align}
the theories are classically equivalent. It is now easy to see that the quantum theories will also be equivalent since the quantization of the Mach conditions imply that the quantum constraints are identical to those of the ADM theory.

\subsubsection{Fixed Endpoints: Unimodular Theory}

In this section we consider the effect of fixing the endpoints of $\tau^0$. According to the definition of background dependence from \scn{BI_BD}, this will introduce a background time. We will, however, not fix a background for the diffeomorphism invariance. Thus, we still have the Mach condition $\zeta^a \approx 0$ for the Mach variation of $\xi_a$.

The constraint algebra is no longer first class after the lifting Mach condition $p_0\approx 0$ because the scalar constraints no longer close on the vector constraints. From (\ref{eq:ham_diff}) and \eq{ham_constraint_ham},
\equa{
  \pb{g^{-1/2} \ham(x)}{\diffadm(y)} = -(g^{-1/2} p_0)^{;a} \delta(x,y),
}
which implies the secondary constraint
\equa{\label{eq:cosmo_constraint}
  -\nabla_a \Lambda = 0,
}
where $\Lambda = -g^{-1/2} p_0$ is the undensitized momentum conjugate to $\tau^0$. The constraint algebra is now first class. Using the Lagrange multipliers $\tau^a$, the total Hamiltonian is
\equa{
  \htot = H_\text{ADM} + 2Np_0 + N_a \zeta^a - \tau^a \nabla_a \Lambda.
}

The secondary constraint \eq{cosmo_constraint} assures that $\Lambda$ is a spatial constant. Given the equations of motion $\dot\tau^0 = N$ and $\dot\Lambda = 0$, one might expect that the $\dot\tau^0\Lambda$ term in the action is analogous to adding a cosmological constant term to the potential. Indeed this is what happens. Since the action is linear in $\zeta^a$, we can integrate out $\zeta^a$ by inserting the equation of motion $\dot\xi_a = N_a$ and the Mach condition $\zeta^a = 0$. This leads to
\begin{multline}
	S_{\text{uni}} = \int d\lambda\, d^nx \, \lf[ \dot g_{ab} \pi^{ab} + \dot\tau p_0 + \sqrt{g} \tau^a\nabla_a \Lambda - \rt. \\
	\lf. N_a\lf( \nabla_b \pi^{(ab)} \rt) - N\lf( \frac{1}{\sqrt{g}} G_{abcd}\pi^{ab}\pi^{cd} - \sqrt{g} (R -  2\Lambda_\text{tot}) \rt) \rt],
\end{multline}
which is identical to the action of unimodular gravity considered by Henneaux and Teitleboim \cite{Henneaux_Teit:unimodular_grav}. Unimodular gravity was originally proposed as a possible solution to the problem of time and was developed extensively in \cite{Unruh:unimodular_grav,Unruh_Wald:unimodular,brown:gr_time}.

Note that $\Lambda_\text{tot} = \Lambda + \Lambda'$. It is the observable value of the cosmological constant. In this context, it will depend on the boundary conditions imposed on the cosmological time
\equa{
  T = \int_{\lambda_\text{in}}^{\lambda_\text{fin}} d\lambda\int_\Sigma d^nx\, \sqrt{g}\, \dot\tau^0.
}
In \cite{Smolin:unimodular_grav}, it is shown that the fact that $\Lambda_\text{tot}$ is an integration constant protects its value against renormalization arguments that predict large values of $\Lambda'$. This provides a possible solution to the cosmological constant problem.

These results show that unimodular gravity is obtained by inserting a background time, according to the definition of background dependence given in this paper, into general relativity. The quantization of this theory is known to lead to a time \emph{dependent} Wheeler-DeWitt equation \cite{brown:gr_time}. This supports the claim that we have inserted a genuine background time. Although there are some hints that unimodular gravity contains unitary cosmological solutions (see \cite{Sorkin:forks_unimodular,sorkin:unmodular_cosmology}), it is clear that unimodular gravity will not be able to solve all problems of time in quantum gravity. As was pointed out by Kucha\v r in \cite{Kuchar:unimodular_grav_critique}, the background time in unimodular gravity is global whereas foliation invariance in general relativity presents several additional challenges. These complications are introduced by the local square root and, therefore, would not have analogues in the finite--dimensional models and the projectable--lapse theories. Furthermore, simply inserting a background should not be thought of as a genuine solution to the problem of time because background dependent theories violate Mach's principle and, I would argue, should not be thought of as fundamental (unless one has other good reasons for believing in an absolute time). Instead, one should think of background dependent theories as having emerged, under special conditions, out of a fundamental background--independent theory.

\section{Outlook}

As has been pointed out throughout this text, in this paper we only consider how to apply best matching to theories where the action is globally gauge invariant. However, in the case where the action has no global symmetries, the best--matching procedure can still be applied. As described in \cite{barbour_el_al:scale_inv_gravity}, the $\omega$'s appear explicitly in the action as well as the $\dot\omega$'s but the generalized rules for Mach variation essentially require that these be treated as independent parameters. Though the $\dot\omega$'s still behave like connections, the $\omega$'s combine with the metric and seem to behave in a way similar to that of Goldstone bosons.\footnote{For a modern review of Goldstone bosons and symmetry breaking, see \cite{Burgess:goldstone_b_review}.} Exploring a possible connection between this more general case of best matching and spontaneous symmetry breaking would be an interesting extension of this work.

After establishing a distinction between BD and BI theories, a natural question to ask is: when do the different theories become important? It may be possible to use effective field theory techniques to determine precisely how BD theories can emerge out of BI ones. Understanding the exact mechanisms for this emergence and the conditions under which it could happen would be vital, for instance, in determining the circumstances under which space and time could emerge out of quantum gravity.


\begin{acknowledgments}
    I would like to sincerely thank Julian Barbour whose best--matching framework and clear thinking on relational ideas have made this work possible. I would also like to thank him for valuable input on the presentation of the text, Fotini Markopoulou for helping to organize sessions on Mach's principle at the Perimeter Institute, and Lee Smolin for having the patience and understanding to give me the freedom to stumble through my own ideas. Research at the Perimeter Institute is supported in part by the Government of Canada through NSERC and by the Province of Ontario through MEDT. I also acknowledge support from an NSERC Postgraduate Scholarship, Mini-Grant MGA-08-008 from the Foundational Questions Institute (fqxi.org), and from the University of Waterloo.
\end{acknowledgments}




\bibliographystyle{utphys}
\bibliography{mach}

\end{document}